\documentclass{article}

\usepackage[final]{neurips_2023}

\usepackage{amsthm}
\usepackage[utf8]{inputenc} 
\usepackage[T1]{fontenc}    
\usepackage{hyperref}       
\usepackage{url}            
\usepackage{booktabs}       
\usepackage{amsfonts}       
\usepackage{nicefrac}       
\usepackage{microtype}      
\usepackage{xcolor}         
\usepackage{graphicx}
\usepackage{amsmath}
\usepackage{wrapfig}

\newcommand{\src}{G}
\newcommand{\tar}{F}
\newcommand{\res}{\hat{F}}

\newcommand{\brwonian}{\mathbf{w}}
\newcommand{\scorenet}{s_\theta}

\newcommand{\datac}{\textit{IXI dataset}}

\usepackage[capitalize]{cleveref}
\crefname{section}{Sec.}{Secs.}
\Crefname{section}{Section}{Sections}
\Crefname{table}{Table}{Tables}
\crefname{table}{Tab.}{Tabs.}
\title{Diffusion based  Zero-shot  
 Medical Image-to-Image Translation for Cross Modality Segmentation}

\author{%
  Zihao Wang \\
    Athinoula A. Martinos Center for Biomedical Imaging \\ Massachusetts General Hospital and Harvard University\\ 02129, Boston, US.
  \And
    Yingyu Yang \\
    Inria centre at Université Côte d'Azur\\
    06902 Valbonne, France.
    \And
Yuzhou Chen\\
    Department of Computer and Information Sciences at Temple University\\ 19122, Philadelphia, US. 
    \And
    Tinting Yuan \\
    Institute of Computer Science at Georg-August-University of Göttingen
    \\ 37073 Göttingen, Germany. 
    \And
    Maxime Sermesant \\
    Inria centre at Université Côte d'Azur\\
    06902 Valbonne, France.  
    \And
    Herv\'{e} Delingette\\
    Inria centre at Université Côte d'Azur\\
    06902 Valbonne, France.
    \And
    Ona Wu\\
    Athinoula A. Martinos Center for Biomedical Imaging \\ Massachusetts General Hospital and Harvard University\\ 02129, Boston, US.
}

\begin{document}
\newtheorem{proper}{Property}
\newtheorem{prop}{Proposition}
\newtheorem{defi}{Definition}
\newtheorem{lemma}{Lemma}
%
\maketitle
Cross-modality image segmentation aims to segment the target modalities using a method designed in the source modality. Deep generative models can translate the target modality images into the source modality, thus enabling cross-modality segmentation.
However, a vast body of existing cross-modality image translation methods relies on supervised learning. In this work, we aim to address the challenge of zero-shot learning-based image translation tasks (extreme scenarios in the target modality is unseen in the training phase).
To leverage generative learning for zero-shot cross-modality image segmentation, we propose a novel unsupervised image translation method. The framework learns to translate the unseen source image to the target modality for image segmentation by leveraging the inherent statistical consistency between different modalities for diffusion guidance. 
Our framework captures identical cross-modality features in the statistical domain, offering diffusion guidance without relying on direct mappings between the source and target domains. This advantage allows our method to adapt to changing source domains without the need for retraining, making it highly practical when sufficient labeled source domain data is not available.
The proposed framework is validated in zero-shot cross-modality image segmentation tasks through empirical comparisons with influential generative models, including adversarial-based and diffusion-based models. 

\section{Introduction}
Leveraging existing tools to handle data across multiple modalities presents both practical benefits and inherent challenges. For example, in the medical imaging field, numerous resources are readily available for image segmentation within certain modalities, such as 3D T1-weighted MRI (referred to as T1w). However, for other modalities like Proton Density-weighted (PDw MRI), there are difficulties. It would be more resource-efficient to employ segmentation tools initially designed for T1w images on PDw images, instead of hunting for extra segmentation tools crafted specifically for the PDw modality. One promising avenue to navigate this issue is zero-shot cross-modality image translation.

Several methods grounded on generative adversarial networks (GAN) \cite{8709985, gan1, gan2, gan3, gan4, gan5, cmtgan7} have been put forth to tackle the image translation challenge. In these techniques, the generative model typically models the destination modality straightaway, thereby ensuring translation authenticity \cite{gan8, gan7}. Notably, designs based on GAN often incorporate intricate adversarial structures and demand the crafting of modality-specific loss functions tailored to diverse translation endeavors \cite{sdeit, WANG2021101990}.
While GAN-based translations can function without matching datasets during the training phase, they still hinge on data from the original domain. This dependence can pose difficulties, especially when there's a scarcity of source domain data, complicating the equilibrium of cycle consistency training \cite{9612090, 9195151}.
This issue becomes even more pronounced when only a scant amount of samples is present for the target modality. 

Recent studies \cite{dhariwal2021diffusion, diff_cond1, diff_con2, nonuni} have unveiled that score-based generative models outperform GAN-based models. Muzaffer \textit{et al.} \cite{GANDIFF} introduced SynDiff, a framework with cycle consistency and dual diffusion for enhanced semantic alignment. Nonetheless, this method entails a doubled computational overhead, necessitates pre-training a generator to assess associated source images, and mandates source domain data, which deviates from the zero-shot learning paradigm. Meanwhile, Meng \textit{et al.} \cite{sdeit} recommended harnessing a Diffusion Model (termed SDEdit) \cite{songsde} to effectuate image translation via zero-shot learning. In contrast with GAN-based models, SDEdit exhibits superiority by its streamlined model configuration and a more straightforward loss function.
However, a limitation of SDEdit emerges in the cross-modality translation context. It is predicted on perturbation-based guidance, which inherently assumes that both the original and destination modalities can be influenced uniformly by noise. Such an assumption often proves invalid in cross-modality image translation, for instance, transitioning from MRI PDw to T1 imaging.

\begin{figure}[t!]
    \centering
    \includegraphics[width =\textwidth]{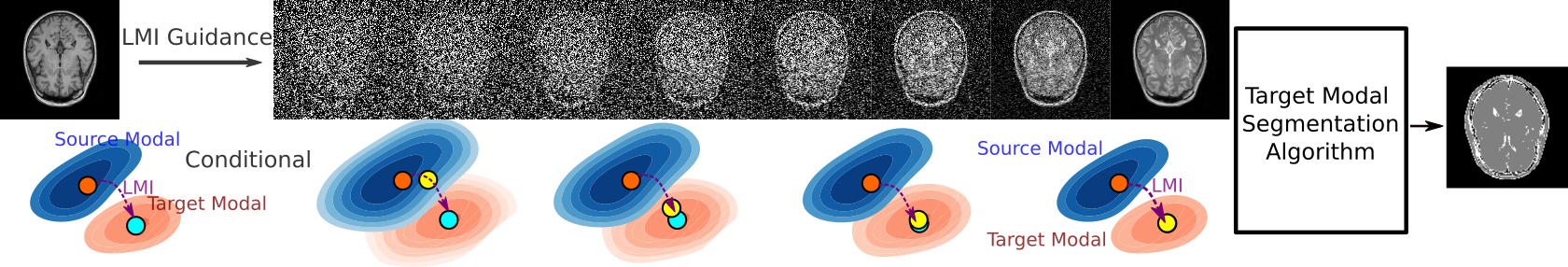}
    \caption{Schematic diagram shows the LMI-guided diffusion for zero-shot cross-modal segmentation. The blue and orange contours are source and target distributions. The blue dot in the orange contour represents the target datapoint of the source datapoint (orange dot in the blue contour) in the source distribution. LMIDiffusion uses explicit statistical features (LMI) to navigate the next step (yellow dot), providing continuous guidance (yellow dot) from start to finish. In the end, the translated image can be segmented using arbitrary segmentation methods that were trained only on the target modality.}
    \label{fig:blation}
\end{figure}

To address the hurdles posed by zero-shot learning in cross-modality translation, our approach leverages statistical feature-wise homogeneity to condition a diffusion model tailored for zero-shot cross-modality image translation.
This strategy leverages a connection between the origin and the target modalities, capitalizing on their localized statistical features for accomplishing cross-modality image transition via the diffusion paradigm. Fig. \ref{fig:blation} illustrates our statistical feature-driven diffusion model ($LMI$: Locale-based Mutual Information), which performs zero-shot image translation for cross-modality image segmentation.
Our proposition is an unsupervised Zero-shot Learning framework capable of navigating translations amidst previously unseen modalities.
\section{Method}
\subsection{Diffusion Model for Cross-modality Image Translation}
The cross-modality image translation can be tackled by score-matching frameworks for generating target data (represented as $\tar$) based on source data $\src$. This is followed by employing a perturbed source domain  $\src$ to condition the step-by-step diffusion process \cite{sdeit, GANDIFF, ozdenizci2022, EnergyGuided, kawar2022denoising, nonuni}.
Yet, when viewed through the lens of zero-shot learning for image translation, the data from the source domain remains inaccessible during the learning phase. However, it is posited that the local statistical attributes between the source and target modalities bear a semblance. Maximizing Mutual Information (MI) has been validated as a potent strategy to equip neural constructs with the capability to model non-linear mappings \cite{hjelm2018learning}. To capture those shared representations for image translation, we propose using MI to model the local statistical features in the denoising process.

\subsection{Local-wise Mutual Information}
To distill semantic information from the dataset for conditioning, it is necessary to translate the raw data into statistical features as encapsulated by MI.
Given an image $X$, for point, $x_i \in X$ at position $i$, the local-wise statistical information at $i$ can be captured through the probability density function (PDF) $p_{\delta_{x_i}}(\cdot)$ of the neighborhood area $\delta_{x_i}$ of $x_i$. 

Concerning other data points, represented as $x_j$, that reside within the neighborhood, $\delta_{x_i}$, of $x_i$, the statistical features can be ascertained via the PDF: $p_{\delta_{x_j}}(\cdot)$, with $j$ being an element of $\delta_i$.
The local-wise MI ($LMI$) from image $X$ to image $Y$ at point $x_i$ is defined through:
\begin{equation}
    \label{eq;LMI}
        LMI_\delta(x_i,y_j) = 
        \sup \iint p_\delta(x, y) \log \frac{p_\delta(x, y)}{p_{\delta_{x_i}}(x)p_{\delta_{y_j}}(y)} dx dy, \forall y_j \in \delta_{x_i}. 
\end{equation}

We employ Eq. \ref{eq;LMI} as a guidance to condition every training step of diffusion. This is realized by evaluating the $LMI(x_0, y_t), t \in [0, T]$ throughout the training steps of the score network.
\begin{proper}
\label{theo;optmum}
The upper bound of the $LMI$ from $X$ to $Y$ at location $i$ is: $LMI_\delta(x_i,y_i) \leq LMI_\delta(x_i,x_i)$
, which is the optimum informative match between $X$ and $Y$ at point $x_i$.
\end{proper}
\begin{proper}
\label{theo;error}
The translation error for the LMIDiffusion generation is $\underset{\Delta LMI \to 0}{\lim}\Delta \mathbb{E}\res=0$.
\end{proper}
In the supplemental section, we prove the properties \ref{theo;optmum} and \ref{theo;error}.
Property \ref{theo;optmum} underscores that $LMI$ attains statistical congruence under the condition:$p_\delta(X) = p_\delta(Y), \forall \delta \in X$. Consequently, throughout the training iterations, the $LMI$ consistently peaks at an identical locale between $X_0$ and $X_t$ in training steps. Yet, when $p_\delta(X) \neq p_\delta(Y)$, the $LMI$ achieves the local maximum at $j$, which is located in the neighborhood $\delta_{y_i}$ of $y_i$. 
The process of MI determination can be expedited by converting iterative mutual information computations into tensor-based operations. Such tensor operations can further benefit from speed enhancements through techniques like memory duplication and parallel reduction when executed on a GPU \cite{cudac}.

\subsection{Conditioning the Diffusion through the LMI}
During the training phase, we can guide the noise perturbation procedure by incorporating the 
 $LMI$ between a data point, denoted as 
$\tar$, and its perturbed counterpart, $\tar_t$ , into the score network, symbolized as 
$\scorenet$. This is realized by adapting the training objective of score matching as:
\begin{equation}
    \arg \min_\theta \mathbb{E}_{t\sim \mathcal{U}(0,T)}
    \{ \mathbb{E}_{x_0 \sim p_0(x)} 
    \mathbb{E}_{x_t \sim q_{\sigma_t}(x_t, x_0)}
    [\frac{1}{2}| \scorenet(\res_t, LMI(\tar; \tar, \tar_t), t) - \dfrac{\partial \log p_{\sigma_t} (x_t|x_0)}{\partial x_t}|^2] \}
    \label{eq;losssdeLMI}
\end{equation}

During the sampling phase, we can integrate the suggested conditioning operator into the SDE:
\begin{multline}
\label{eq;embedded}
    d\res_t =- \frac{d\sigma_t^2}{dt} \scorenet(\res_t, LMI(\src; \src, \res_t), t) dt + 
    \sqrt{\frac{d\sigma_t^2}{dt}} d\brwonian_t, ~~~~t_{T \rightarrow 0} \in [0, T]
\end{multline}
and use a naive Euler-Maruyama solver to integrate the SDE. The sampled images can be segmented directly by the tool designed solely for the target domain image segmentation.
\section{Experiment and Result}
\begin{figure}
    \centering
    \vspace{-1mm}
    \includegraphics[width=\textwidth]{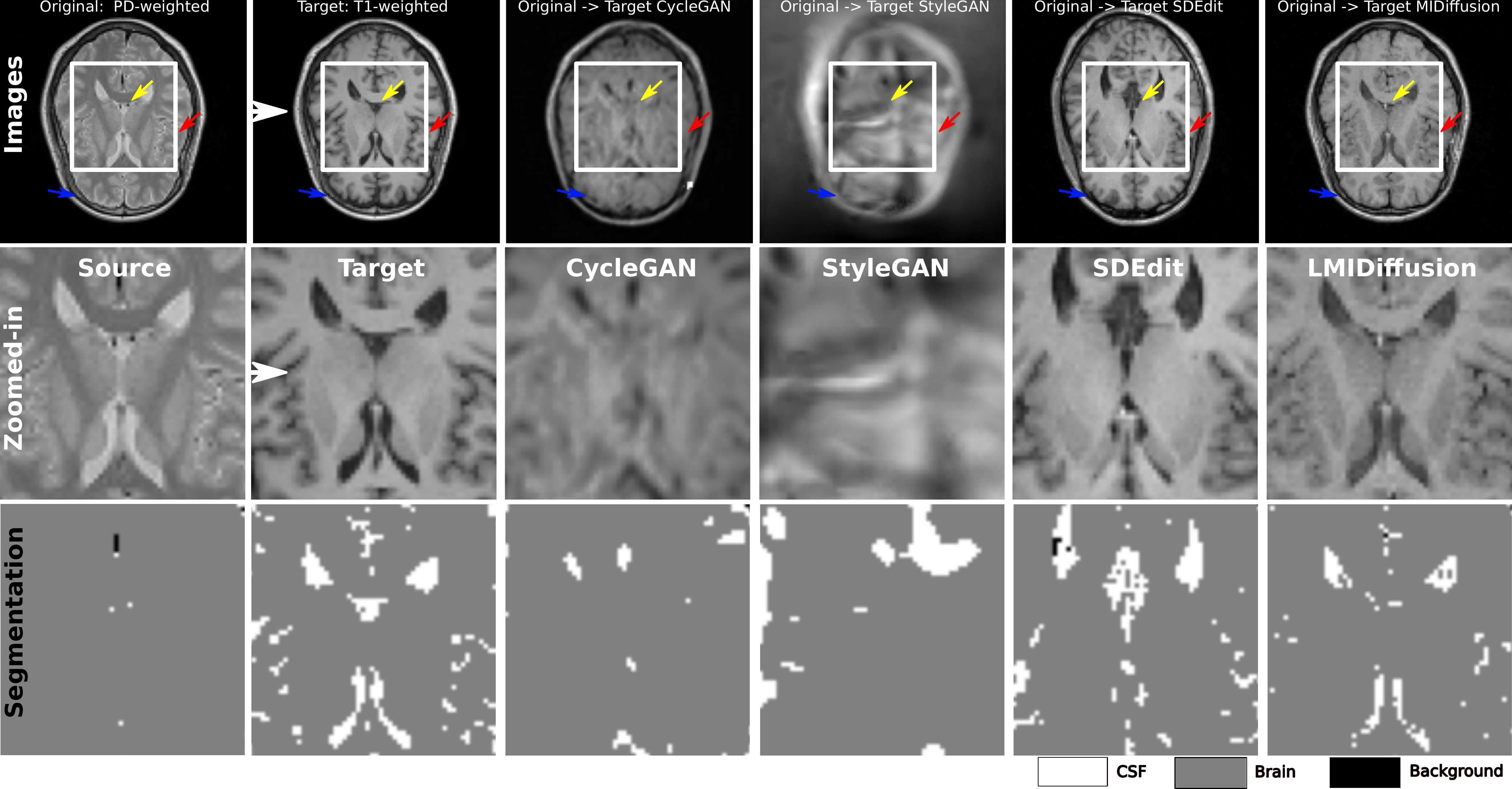}
    \vspace{-6mm}
    \caption{Qualitative evaluation of different models' translation results. The first two rows show target and original modality images, with close-ups of ROIs, followed by transformations from CycleGAN, StyleGAN, SDEdit, and LMIDiffusion. The subsequent row displays binarized segmentation results in the ROIs using a 3 clusters K-Means method for segmentation, trained solely on the target modality.
}
    \label{fig;segmentation}
    \vspace{-5mm}
\end{figure}
\paragraph{Dataset} we utilize the \datac~dataset \cite{marcoixi} to demonstrate the efficacy of the proposed cross-modality method for image segmentation. This dataset comprises 600 pre-aligned multi-modality images from healthy subjects. We undertake translation tasks between PD and T1w modalities. From a subset of the \datac~, a training set consisting of 300 slices (drawn from 100 subjects) and a testing set consisting of 75 slices (drawn from 25 subjects) were formulated.
\paragraph{Experiment} We compare our LMIDiffusion (\textbf{unsupervised}) with a few-shot learning-based CycleGAN (\textbf{supervised}) translation model \cite{CycleGAN2017}, a GAN inversion-based approach \cite{styleGAN2} (\textbf{unsupervised}) and a diffusion-based method SDEdit \cite{sdeit} (zero-shot \textbf{supervised} learning with SDE-based perturbing guidance). 
The CycleGAN (\textbf{supervised}) in this experiment will be allowed to see both the full target domain dataset and a small group (about $11\%$ of dataset \datac) of the source domain dataset. 
The second baseline is a GAN inversion-based approach (\textbf{unsupervised}). A StyleGAN2-ADA \cite{styleGAN2} is allowed to see the target domain training data. The out-domain guided generation is performed through 5000 steps of optimization of inversion in the latent space of the trained StyleGAN2-ADA \cite{ganinversion}. An extra network is introduced in the generation step for inversion. The SDEdit \cite{sdeit} (\textbf{unsupervised}) uses a distribution perturbation guidance for diffusion. 
Our score network structure is the same as the UNet-like score matching network\cite{sdeit}, which was optimized through an Adam optimizer with a learning rate of $3e-4$. The model was trained on an NVIDIA DGX Station with four Tesla V100 GPUs, over 300K iterations. For the segmentation backend, we employed the K-Means (5 clusters) algorithm, which was trained exclusively on the target modality. Subsequent segmentation was performed on the translation results obtained from the various methodologies mentioned above.

\paragraph{Result} Fig.~\ref{fig;segmentation} shows both the translation and segmentation outcomes from various models. The bottom rows provide a clear illustration of segmentation results based on translated images. Notably, the segmentation derived from LMIDiffusion translations closely mirrors the segmentation ground truth. While direct segmentation using a method trained on the target image is not feasible for the source domain image, our translation facilitates this, yielding commendable segmentation results on the translated image. Our approach not only offers maximal resemblance to the translation target (PDw) but also retains the superior anatomical features of the original modality (T1w). Although the CycleGAN is trained with supervised (few-shot) data, it fails to produce high-quality images. This shortcoming can be attributed to GAN-based models' difficulty in discerning relationships between source and target modalities when training data from both domains is limited. On the other hand, while SDEdit does produce images in the PDw domain, it compromises the anatomical features of the original domain, rendering it less effective for zero-shot image segmentation.

We present the Dice score, PSNR and SSIM values computed based on different models in Table \ref{tab:your_label_here}. It is evident that the LMIDiffusion achieves a leading average Dice score of $0.88\pm0.05$. This is closely followed by the CycleGAN-based method with a score of $0.85\pm0.07$. The SDEdit technique demonstrates suboptimal performance with a Dice score of $0.82$. In contrast, the GAN-inversion method (StyleGAN) yields a result of $0.66$, emphasizing its challenges in cross-modality image translation.
In terms of translation quality metrics, LMIDiffusion leads the pack, boasting PSNR and SSIM values of $20.22$and SSIM of $0.69$, respectively. CycleGAN follows suit with a PSNR of $18.88$ and SSIM of $0.27$, while SDEdit reports a PSNR of
$15.96$ and SSIM of $0.50$. The StyleGAN method, consistent with its lower Dice score, presents the lowest PSNR of
$10.15$ and SSIM of $0.06$, further evidences the limitation of the GAN-inversion approach for zero-shot image translation task.

\begin{table}
\centering
\caption{Quantitative evaluation of the translation performance for different methods.}
\label{tab:your_label_here}
\begin{tabular}{ccccc}
PDw$\rightarrow$T1          & CycleGAN\cite{CycleGAN2017}           & StyleGAN\cite{styleGAN2}          & SDEdit \cite{sdeit}          & LMIDiffusion           \\ 
\hline
Dice Score & 0.85 $\pm$ 0.07  & 0.66 $\pm$ 0.10  & 0.82 $\pm$ 0.06  & \textbf{0.88} $\pm$ 0.05   \\
PSNR       & 18.88 $\pm$ 1.26 & 10.15 $\pm$ 1.49 & 15.96 $\pm$ 1.54 & \textbf{20.22} $\pm$ 1.43  \\
SSIM       & 0.27 $\pm$ 0.04  & 0.06 $\pm$ 0.03  & 0.50 $\pm$ 0.06  & \textbf{0.69} $\pm$ 0.06   \\
\hline\hline
\end{tabular}
\vspace{-5mm}
\end{table}

\section{Conclusion}
We propose a novel method for zero-shot cross-modality image translation with application in cross-modality image segmentation. For challenging of zero-shot learning, our method outperforms existing GAN-based and diffusion-based methods regarding translation quality and zero-shot segmentation performance.  The results show that the proposed LMI-guided diffusion is a promising approach for cross-modality image translation-based segmentation in a zero-shot learning way. The performance of the model can be potentially improved by introducing one-shot or few-shot learning to refine the LMI computing for diffusion guidance.
\appendix
\section{Appendix}
\label{appendixb}
\begin{proof}[Property \ref{theo;optmum}]
\begin{align}
     &LMI_\delta(x_i,y_j)= \sup \iint p_\delta(x, y) \log \frac{p_\delta(x, y)}{p_{\delta_{x_i}}(x)p_{\delta_{y_j}}(y)} dx dy \\
    &= \sup \Biggl(\iint p_\delta(x, y) \log \frac{p_\delta(x, y)}{p_{\delta_{y_j}}(y)} dx dy  - 
     \iint p_\delta(x, y) \log{p_{\delta_{x_i}}(x)} dx dy 
    \Biggl) , \forall y_j \in \delta_{x_i} 
\end{align}
Applying Bayes' theorem we obtain:
\begin{align}
    &= \sup \Biggl(\int p_{\delta_{y_j}}(y) 
    \int p_{\delta_{x_i|y_j}}(x|y) \log p_{\delta_{x_i|y_j}}(x|y)dx dy \\
    &- \iint p_\delta(x, y) \log{p_{\delta_{x_i}}(x)}dxdy \Biggl) , \forall y_j \in \delta_{x_i}  \\
    &= \sup (h_{\delta_{x_i}}(x)-h_{\delta{x_i|y_j}}(x|y)), \forall y_j \in \delta_{x_i}  \\
    &\leq \sup (h_{\delta_{x_i}}(x)) , \forall y_j \in \delta_{x_i}  
\end{align}
where $h$ is entropy that measures the informative of the random variables.
We can have the following corollaries:

\textbf{LMI bound of Forward SDE}: If and only if $x_i = y_j$: $LMI_\delta(x_i,x_i) = h_{\delta_{x_i}}(x) - h_{\delta_{x_i|x_i}}(x|x)) =h_{\delta_{x_i}}(x)$, which is the upper bound; therefore for the forward SDE: $LMI_\delta(x_i,y_j) \leq LMI_\delta(x_i, x_i), \forall y_j \in \delta_{x_i}$.

\textbf{LMI bound of Reverse SDE}:
If $x_i \neq y_j$, $x_i \neq z_j$: $LMI_\delta(x_i,y_j) = h_{\delta_{x_i}}(x) - h_{\delta_{x_i|y_j}}(x|y)$, then the $LMI_\delta(x_i,y_j)$ is the supremum in $\delta$ neighborhood, bounded by $LMI_\delta(x_i, x_i)$: $LMI_\delta(x_i, z_j) \leq LMI_\delta(x_i,y_j) < LMI_\delta(x_i, x_i), \forall y_j, z_j \in \delta_{x_i}$
\end{proof}
\begin{proof}[Property \ref{theo;error}]
For cross-modality data translation, the reverse SDE for$ t_{T \rightarrow 0} \in [0, T]$ is:
\begin{equation}
\label{eq;revsde}
    d\res_t =- \frac{d\sigma_t^2}{dt} \scorenet(\res_t, LMI(\src; \src, \res_t), t) dt + \sqrt{\frac{d\sigma_t^2}{dt}} d\brwonian_t
\end{equation}
Its solution is:
\begin{equation}
\label{eq;solution}
    \res_t = \res_0 - \int_0^t \frac{d\sigma_s^2}{ds} \scorenet(\res_s, LMI(\src; \src, \res_s), s) ds +
    \int_0^t\sqrt{\frac{d\sigma_s^2}{ds}} d\brwonian_s
\end{equation}
Take the expectation on both side of Eq.~\ref{eq;solution}:
\begin{multline}
\mathbb{E}\res_t = \mathbb{E}\res_0 - \mathbb{E}\int_0^t \frac{d\sigma_s^2}{ds} \scorenet(\res_s, LMI(\src; \src, \res_s), s) ds +
\mathbb{E}\int_0^t\sqrt{\frac{d\sigma_s^2}{ds}} d\brwonian_s\\
= \res_0 - \int_0^t \mathbb{E}\frac{d\sigma_s^2}{ds} \scorenet(\res_s, LMI(\src; \src, \res_s), s) ds
\end{multline}
Thus, the expectation of generation error is:
\begin{align}
\label{eq;error}
\Delta \mathbb{E}\res_t = - \int_0^t \frac{d\sigma_s^2}{ds}\mathbb{E}\Biggl[\scorenet(\res_s, LMI(\src; \src, \res_s), s) - 
\scorenet(\res_s, LMI(\tar; \tar, \res_s), s)\Biggl] ds
\end{align}
Eq. \ref{eq;error} shows that the expectation of generation error is the accumulated expectation of the conditioning error of score model $\scorenet$ between the training and testing steps.

Without loss of generality, assuming that the score model $\scorenet$ satisfied \textit{additivity} and \textit{homogeneity} for the guidance $LMI$:
\begin{align}
\label{eq;linear}
\scorenet(\res_t, LMI(\cdot, t), t) = \alpha(\res_t, t) LMI(\cdot, t) + \beta(\res_t, t)
\end{align}
Substituting Eq. \ref{eq;linear} into Eq. \ref{eq;error}:
\begin{align}
\label{eq;error2}
\Delta \mathbb{E}\res_s &= - \int_0^t \frac{d\sigma_s^2}{ds}\mathbb{E}\Biggl[\alpha(\res_s, s) LMI(\src; \src, \res_s) + \beta(\res_s, s) - \\
&\alpha(\res_s, s) LMI(\tar; \tar, \res_s) - \beta(\res_s, s) \Biggl] ds\\
&= - \int_0^t \frac{d\sigma_s^2}{ds}\alpha(\res_s, s)\mathbb{E}\Biggl[LMI(\src; \src, \res_s) - 
LMI(\tar; \tar, \res_s) \Biggl] ds
\end{align}
Denote: $\Delta LMI = LMI(\src; \src, \res_s) - LMI(\tar; \tar, \res_s) $, the error for the expectation of $LMI$-guided generation is:
\begin{align}
\label{eq;error3}
\Delta \mathbb{E}\res_s &= - \int_0^t \frac{d\sigma_s^2}{ds}\alpha(\res_s, s)\Delta LMI ds
\end{align}
Thus,
$\underset{\Delta LMI \to 0}{\lim}\Delta \mathbb{E}\res_s=0$
\end{proof}

\bibliographystyle{plain}
\bibliography{ref}

\end{document}